\documentclass{acmsmall}

\pdfoutput=1
\usepackage{graphicx}
\usepackage{rotating}
\usepackage{array}
\usepackage{multirow}
\usepackage{caption}
\setcopyright{rightsretained}

\begin{document}
\sloppy








\title{Autonomy and Reliability of Continuous Active Learning for Technology-Assisted
Review}

\author{GORDON V. CORMACK\affil{University of Waterloo}MAURA R. GROSSMAN\affil{Wachtell, Lipton, Rosen \& Katz*}}
\begin{abstract}
We enhance the autonomy of the continuous active learning method shown by Cormack and Grossman (SIGIR 2014) to be effective for technology-assisted review, in which documents from a collection are retrieved and reviewed, using relevance feedback, until substantially all of the relevant documents have been reviewed. Autonomy is enhanced through the elimination of topic-specific and dataset-specific tuning parameters, so that the sole input required by the user is, at the outset, a short query, topic description, or single relevant document; and, throughout the review, ongoing relevance assessments of the retrieved documents. We show that our enhancements consistently yield superior results to Cormack and Grossman's version of continuous active learning, and other methods, not only on average, but on the vast majority of topics from four separate sets of tasks: the legal datasets examined by Cormack and Grossman, the Reuters RCV1-v2 subject categories, the TREC 6 AdHoc task, and the construction of the TREC 2002 filtering test collection.
\end{abstract}
\begin{bottomstuff}
* The views expressed herein are solely those of the author and should not be attributed to her firm or its clients.
\end{bottomstuff}

\maketitle

\section{Introduction}

Technology-assisted review (\textquotedblleft{}TAR\textquotedblright{})
involves the iterative retrieval and review of documents from a collection
until a substantial majority (or ``all'') of the relevant documents
have been reviewed. Applications include electronic discovery (``eDiscovery'')
in legal matters \cite{Cormack:2014:EMP:2600428.2609601}, systematic
review in evidence-based medicine \cite{lefebvre2008searching}, and
the creation of test collections for information retrieval (``IR'')
evaluation \cite{sand04}. In contrast to ad hoc search, the information
need is satisfied only when virtually all of the relevant documents
have been discovered; as a consequence, a substantial number of documents
are typically examined for each review task. The reviewer is typically
expert in the subject matter, not in IR or data mining. In certain
circumstances, it may be undesirable to trust the completeness of
the review to the skill of the user, whether expert or not. In eDiscovery,
the review is typically conducted in an adversarial context, which
may offer the reviewer limited incentive to conduct the best possible
search. In systematic review, meta-analysis affords valid statistical
conclusions only if the selection of studies for inclusion is reasonably
complete and free of researcher bias. The creation of test collections
is subject to similar constraints: The assessors are not necessarily
search experts, and the resulting relevance assessments must be reasonably
complete and free of selection bias.

For the reasons stated above, it may be desirable to limit discretionary
choices in the selection of search tools, tuning parameters, and search
strategy. Obviating such choices presents a challenge because, typically,
both the topic and the collection are novel for each task to which
TAR is applied, and may vary substantially in subject matter, content,
and richness. Any topic- or collection-specific choices, such as parameter
tuning or search queries, must either be fixed in advance, or determined
autonomously by the review tool. Our goal is to fully automate these
choices, so that the only input required from the reviewer is, at
the outset, a short query, topic description, or single relevant document,
followed by an assessment of relevance for each document, as it is
retrieved.

At the same time, it is necessary for each TAR task to enjoy a high
probability of success. A lawyer engaged in eDiscovery in litigation,
or a researcher conducting a meta-analysis or building a test collection,
is unlikely to be consoled by the fact that the tool works well on
average, if it fails for the particular task at hand. Accordingly,
it is necessary to show that such failures are rare, and that such
rare failures are readily apparent, so that remedial actions may promptly
be taken.

The literature reports a number of search efforts aimed at achieving
high recall, particularly within the context of eDiscovery and IR
evaluation. Most of these efforts require extensive intervention by
search experts, or prior topic- or dataset-specific training, contrary
to our objective. Many search and categorization methods are unreliable,
in that they fail to achieve reasonable effectiveness for a substantial
number of topics, although, perhaps, achieving acceptable effectiveness
on average.

Among approaches that meet our criterion of autonomy, the continuous
active learning (``CAL'') method, and its implementation in Cormack
and Grossman's TAR Evaluation Toolkit (``Toolkit'') \cite{Cormack:2014:EMP:2600428.2609601},
appears to be the standard to beat.%
\footnote{With honorable mention to the interactive relevance feedback (``IRF'')
technique pioneered by Soboroff and Robertson \cite{Soboroff:2003:BFT:860435.860481},
which is quite similar.%
} Yet uncertainties remain regarding its sensitivity to the choice
of ``seed query'' required at the outset, its applicability to topics
and datasets with higher or lower richness, its algorithmic running
time for large datasets, its effectiveness relative to non-autonomous
approaches, and its generalizability to domains beyond eDiscovery.

We hypothesized the impact of various engineering choices with respect
to these considerations and designed an autonomous TAR configuration
(``Auto TAR'') that, assuming our hypotheses to be correct, would
exhibit greater autonomy, superior effectiveness, increased generalizability,
and fewer, more easily detectable failures, relative to existing TAR
methods.

This paper offers a historical review of research efforts to achieve
high recall, followed by a description of our final design and the
results we achieved applying it to the four topics and dataset supplied
in Cormack and Grossman's Toolkit, which we used to pilot the development
of Auto TAR. We then describe the hypotheses underlying our choices,
along with our rationale. Next, we describe our experiments to evaluate
the effectiveness of Auto TAR and competing approaches, relative to
strong baselines; specifically:
\begin{enumerate}
\item The four actual legal matters and datasets studied by Cormack and
Grossman, but not included in the Toolkit, relative to the CAL implementation
in the Toolkit;
\item The 103 subjects of the Reuters RVCV1-v2 dataset, relative to the
CAL implementation in the Toolkit;
\item The 103 subjects of the Reuters RCV1-v2 dataset, relative to the best
text categorization method reported by its authors; 
\item The 50 topics of the TREC 6 AdHoc task, relative to the best-performing
manual efforts reported at TREC; and,
\item The 50 topics of the TREC 2002 Filtering Track, relative to the labeling
effort conducted by the Track coordinators.
\end{enumerate}
With few exceptions, Auto TAR yields comparable or superior results
to the chosen baselines, when evaluated for all combinations of topic
and representative recall values from 0.1 through 0.9.

We conclude with a brief discussion of the strengths and limitations
of our results, and open questions for further study.

\section{Related Work}

\label{sec:Related-Work}The TAR problem, along with methods to address
it and to evaluate the effectiveness of those methods, resembles but
may be distinguished from well-studied problems in ad hoc retrieval,
relevance feedback, routing and filtering, text categorization, and
active learning.

Blair and Maron \cite{Blair85a} evaluated an iterative search effort
in which skilled paralegals and lawyers collaborated in an interactive
search effort aimed at finding ``all and only the relevant items.''
Realizing that perfection was an impossible standard to meet, the
lawyers ``stipulated that they must be able to retrieve at least
75 percent of all the documents relevant to a given request for information.''
For each of 51 information requests, the reviewers composed Boolean
search queries and assessed the results until they believed they had
seen at least 75\% of the relevant documents ($recall\ge0.75$). In
fact, according to Blair and Maron's evaluation effort, the reviewers
had found, on average, 20\% ($recall\approx0.20$). On the other hand,
it can be deduced from the raw data presented by Blair and Maron,
that about 80\% of the retrieved documents were relevant ($precision\approx0.80$).

Blair and Maron themselves conducted a high-recall search for the
purpose of evaluating the recall of the reviewers' efforts \cite{Blair:1996:SRT:231880.231885}.
Using complete conjunctive normal form (``CCNF'') of semantically
expanded Boolean queries, Blair and Maron identified and sampled subsets
of the collection that were rich in relevant documents. While we do
not know precisely the recall or precision of the sets of documents
identified by these systematic searches, we do know that recall was
five times greater than that of the paralegal and lawyer searches,
while precision was, presumably, many times lower.

To our knowledge, Blair and Maron's method has not been widely used
to construct labeled datasets for IR evaluation. The most common approach
is perhaps the pooling method, used extensively at TREC \cite{voorhees2000vrj}.
The pooling method shares with TAR the objective of identifying substantially
all relevant documents in a collection. To this end, the top-ranked
documents from a large set of independent ranked-retrieval efforts
are combined to form a judging pool, and assessed for relevance. These
relevance assessments (``qrels'') are used to evaluate the retrieval
efforts used to form the pool, and as part of an archival test collection,
to evaluate future efforts. It is well known that the set of relevance
assessments is incomplete, as some relevant documents are excluded
from the pool \cite{zobel98}. However, the literature suggests that,
notwithstanding this incompleteness, the pooling method affords stable
evaluation of IR effectiveness \cite{voorhees2000vrj}.

Although the objective of ad hoc retrieval is typically to produce,
after limited interaction with a search engine, a set or ranked list
of documents for off-line review, a team from the University of Waterloo,
as participants in the TREC 6 AdHoc Task \cite{cormackpalmer98},
conducted a TAR process in which ``the aim of the searchers was to
find and judge as many relevant documents as possible.'' The Waterloo
team spent an average of 2.1 hours per topic composing search queries,
examining the top-ranked results, labeling each document as ``relevant,''
``not relevant,'' or ``iffy.'' After examining 13,064 documents
(an average of 261 documents per topic), the team achieved per-topic
(macro-averaged) recall of 0.8. Because the submissions were padded
to 1,000 documents, it is not possible to deduce per-topic precision
from the archived TREC 6 results; however, we note that the Waterloo
team found 3,058 of 4,611 relevant documents overall, for a micro-averaged
recall of 0.66 and a precision of 0.23.

Independent studies by Voorhees \cite{voorhees2000vrj} and the Waterloo
team \cite{cormackpalmer98} indicate that Waterloo's technique --
interactive search and judging (``ISJ'') -- yields a judging pool
about as effective for IR evaluation as the five-times-larger pool
derived using the pooling method that was employed for the official
TREC 6 evaluation. The ISJ pool, labeled using the official TREC assessors'
judgments, achieved the Kendal rank correlation with respect to the
official mean average precision (``MAP'') results of $\tau\approx0.98$.
Labeled using the Waterloo assessor's judgments, ISJ achieved $\tau\approx0.90$.,
characterized by Voorhees as ``essentially the same comparative evaluation
results.'' Sanderson and Joho \cite{sand04} simulated ISJ using
input from a wide variety of manual search efforts and concluded that,
``ISJ is broadly applicable regardless of retrieval system used or
people employed to conduct the searching process.'' At the same time,
they note that a few ISJ efforts did not fare well, observing that
the TREC runs on which these efforts were based had very low MAP scores,
concluding that, ``it would be unlikely that someone using the Cormack
et al. ISJ method would create such runs, as consistent poor performance
would be noticed by the experimenter.''

Soboroff and Robertson \cite{Soboroff:2003:BFT:860435.860481}, in
constructing qrels for the TREC 2002 Filtering Track, were unable
to use the pooling method, because the qrels were required in advance
for the purpose of evaluating on-line filtering systems. They eschewed
ISJ -- on the grounds that the TREC assessors were subject-matter
experts, but not search experts -- in favor of a relevance feedback
approach. For each of 100 candidate topics, ad hoc search was used
to retrieve and label 100 potentially relevant documents. These labeled
documents were given as relevance feedback to variants of four retrieval
systems, and the top-ranked documents (according to the CombMNZ fusion
method) were selected for labeling by TREC assessors. These labeled
documents were again provided as relevance feedback to the retrieval
systems, and a new batch of top-ranked documents were selected for
review. The process continued for each topic until relatively few
relevant documents were returned, or the assessment budget was exhausted.
Based on these results, the ``best'' 50 topics were selected for
use in the TREC 2002 Filtering Track. Once participants submitted
their search results, the pooling method was used to augment the qrels
for these topics. Robertson and Soboroff report a Kendall rank correlation
of $\tau=0.91$ between MAP results for the Routing Task derived from
the initial and augmented qrels. Sanderson and Joho's subsequent simulations
indicate that relevance feedback, while effective for this purpose,
is not as effective as ISJ.

While ad hoc search involves finding documents relevant to a new topic
in an existing collection, routing (along with text categorization
and batch filtering) involves finding documents relevant to a standing
information need in a new collection \cite{Soboroff:2003:BFT:860435.860481,sebastiani02machine}.
The objective of routing is to rank the collection for subsequent
review, while the objective of text categorization and batch filtering
is to identify the set of responsive documents in the collection without
further review. An existing set of labeled data -- the training set
-- is available for tuning and model construction, while a separate
set of labeled data -- the test set -- is used for evaluation. Machine-learning
methods, such as support vector machines (``SVM''), logistic regression,
and boosting, can, given a large enough training set, generally achieve
both high precision and high recall on the test set. The premier test
collection for such tasks is perhaps the Reuters RCV1-v2 dataset,
for which Lewis \cite{lewis2004rnb} reports that SVMlight, applied
to the tf-idf representation and training/test sets supplied with
the dataset, achieves a macro-averaged score of $F_{1}=0.619$, indicating
that, on average, this method achieves reasonably high recall as well
as reasonably high precision. Nonetheless, there are two topics%
\footnote{For a third topic, there are no relevant documents in the test set;
for this topic, $F_{1}$ is indeterminate.%
} for which there are no relevant examples in the training set, and
the method fails entirely, achieving $F_{1}\approx0.00$.

In active learning \cite{settles2010active}, the learning method
interactively chooses only a subset of the documents from the training
set to be labeled, thereby reducing labeling effort. Lewis and Gale
\cite{lewis1994sequential} compare uncertainty sampling, in which
the hardest-to-classify documents are selected for labeling, with
relevance sampling, in which the most-likely relevant examples are
selected, concluding that uncertainty sampling generally works better
when the proportion of relevant documents in the collection is high,
while relevance sampling works well when the proportion is low. Active
learning methods are the subject of much current research interest;
derivatives of uncertainty sampling and query-by-committee are generally
considered to be the most effective \cite{settles2010active}.

Drucker \cite{drucker2002support} shows that machine-learning methods
are also useful for relevance feedback in ad hoc retrieval. In contrast
to uncertainty sampling and query-by-committee, relevance feedback
focuses on retrieving and labeling documents that are most likely
to be relevant. Assuming that a preliminary query has retrieved ten
documents, of which at least one is relevant and at least one is non-relevant,
Drucker shows that iteratively using an SVM to retrieve and review
several batches of ten documents can achieve very high relative precision.%
\footnote{``Relative precision,'' implemented in trec\_eval\\
{[}\texttt{http://trec.nist.gov/trec\_eval/}{]}, is equal to\\
$\frac{R_{k}}{\min(k,R_{\infty})}$, where $k$ is the number of documents
retrieved, $R_{k}$ is the number of relevant documents retrieved,
and $R_{\infty}$ is the number of relevant documents in the collection.
Relative precision is equal to precision when the number of retrieved
documents is less than the number of relevant documents in the collection,
and equal to recall when the number of retrieved documents is greater.%
} The ability to achieve high recall is not considered.

Within the context of the TREC Legal Track Interactive Task \cite{oard2008overview,Hedin09a,cormack2010overview,grossman2011overview},
a number of participating teams achieved high recall using different
techniques. In 2008 and 2009, Hogan et al. \cite{hogan2008h5} achieved
superior results ($F_{1}=\{0.71,0.80\}$) on two topics using a rule-based
approach: composing myriad Boolean queries, sampling the results,
and, with the aid of a professional linguist, augmenting the queries,
until a high level of recall and precision were achieved. In 2009,
Cormack and Mojdeh \cite{cormack2009machine} used a combination of
ISJ and relevance feedback via logistic regression to achieve superior
results on four topics ($F_{1}=\{0.84,0.76,0.77,0.83\}$). In 2010,
a team from an eDiscovery provider achieved superior results ($F_{1}=0.67$)
on one topic, using a proprietary multi-stage approach -- referred
to as ``predictive coding'' in the legal industry -- that involved
isolating from the dataset a hold-out ``control set'' for tuning
and validation, and a set used to train an SVM using uncertainty sampling.
In 2011, a team from a different eDiscovery provider achieved superior
results ($F_{1}=\{0.54,0.53,0.43\}$), using a different proprietary
implementation of predictive coding.

Cormack and Grossman \cite{Cormack:2014:EMP:2600428.2609601} provide
a taxonomy of TAR systems with three major categories: ``continuous
active learning'' (``CAL''); ``simple active learning'' (``SAL'');
and ``simple passive learning'' (``SPL''). Each starts with an
initial training set (``seed set'') which may be selected randomly
or by searching. CAL abstracts the technique of Cormack and Mojdeh
\cite{cormack2009machine}, using search only to identify the seed
set, and using exclusively relevance feedback thereafter. SAL abstracts
the predictive coding method outlined above, in which a classifier
is trained using supervised active learning, and then applied to the
entire collection. SPL abstracts a more primitive version of predictive
coding, in which the entire training set is selected either randomly
or by searching, without the aid of active learning. 

\begin{table}
\begin{centering}
\begin{tabular}{|c|cc|cc|}
\hline 
\multirow{2}{*}{Topic} & \multicolumn{2}{c|}{Cormack and Mojdeh} & \multicolumn{2}{c|}{CAL}\tabularnewline
 & Recall & Effort & Recall & Effort\tabularnewline
\hline 
201 & 0.78 & 6,145 & 0.75 & 6,000\tabularnewline
202 & 0.67 & 12,624 & 0.75 & 11,000\tabularnewline
203 & 0.87 & 4,369 & 0.75 & 6,000\tabularnewline
207 & 0.76 & 34,446 & 0.75 & 11,000\tabularnewline
\hline 
\end{tabular}
\par\end{centering}

\caption{\label{tab:Comparison-of-Cormack}Recall effort of Cormack and Mojdeh
vs. CAL.}
\end{table}
Cormack and Grossman also provide a toolkit for evaluating TAR systems,
and, using that toolkit, compare the effectiveness of their implementations
of CAL, SAL, and SPL. They concluded that CAL (using a simple ``seed
query'' to identify the seed set) yielded generally superior results
to SAL and SPL, although SAL -- given an oracle to determine the optimal
``stabilization'' point -- could match CAL for a specific target
recall level \cite{Cormack:2014:EMP:2600428.2609601}. Table \ref{tab:Comparison-of-Cormack}
juxtaposes Cormack and Grossman's ``75\% recall effort'' values
-- the number of documents that must be reviewed (counting both those
for training and for any subsequent review) to achieve $recall=0.75$
-- with the effort expended by Cormack and Mojdeh \cite{cormack2009machine}
to achieve similar results. While uncontrolled differences between
the two studies preclude direct comparison, we note that they are
of the same order.

\section{Autonomous TAR}

\label{sec:Autonomous-TAR}

\subsection{The method}

\begin{figure*}
\includegraphics[width=1\columnwidth]{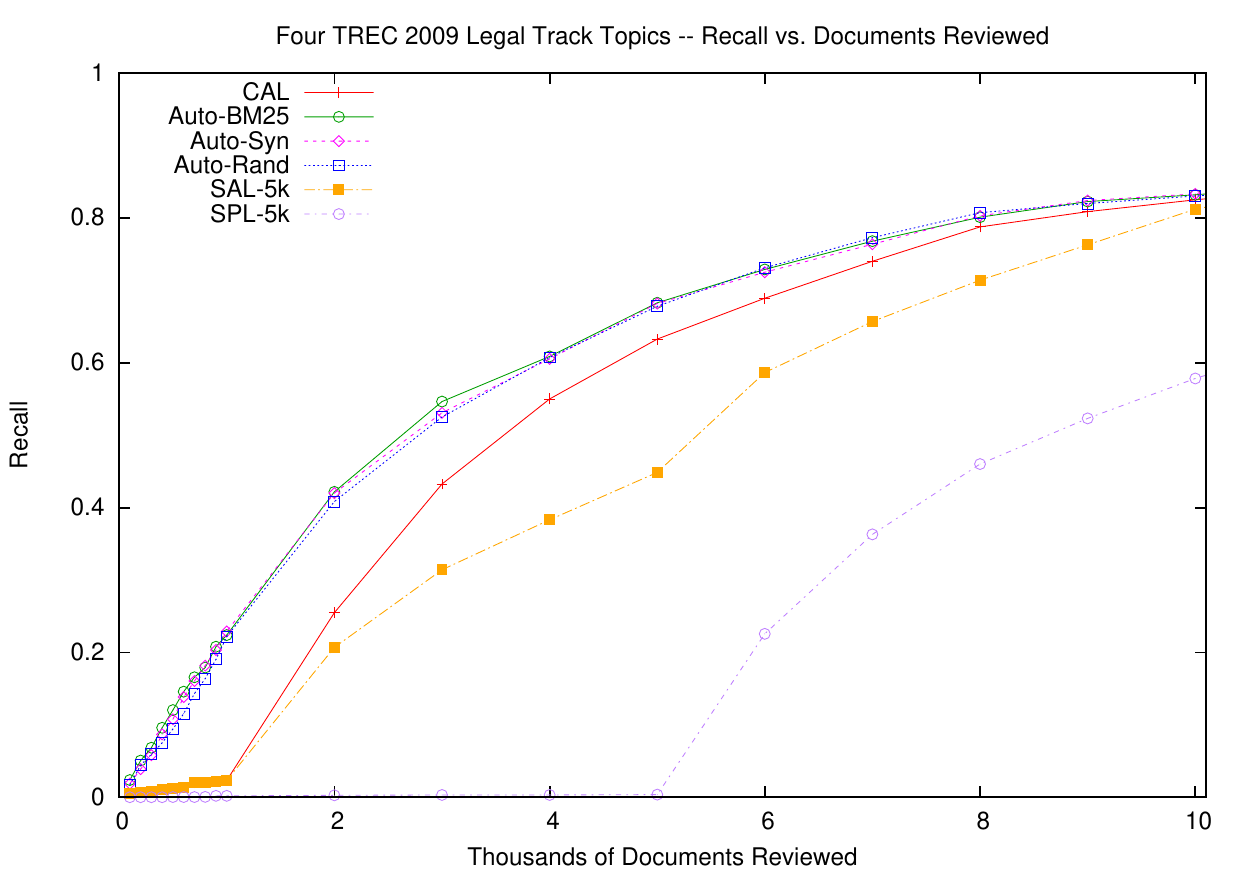}

\includegraphics[width=1\columnwidth]{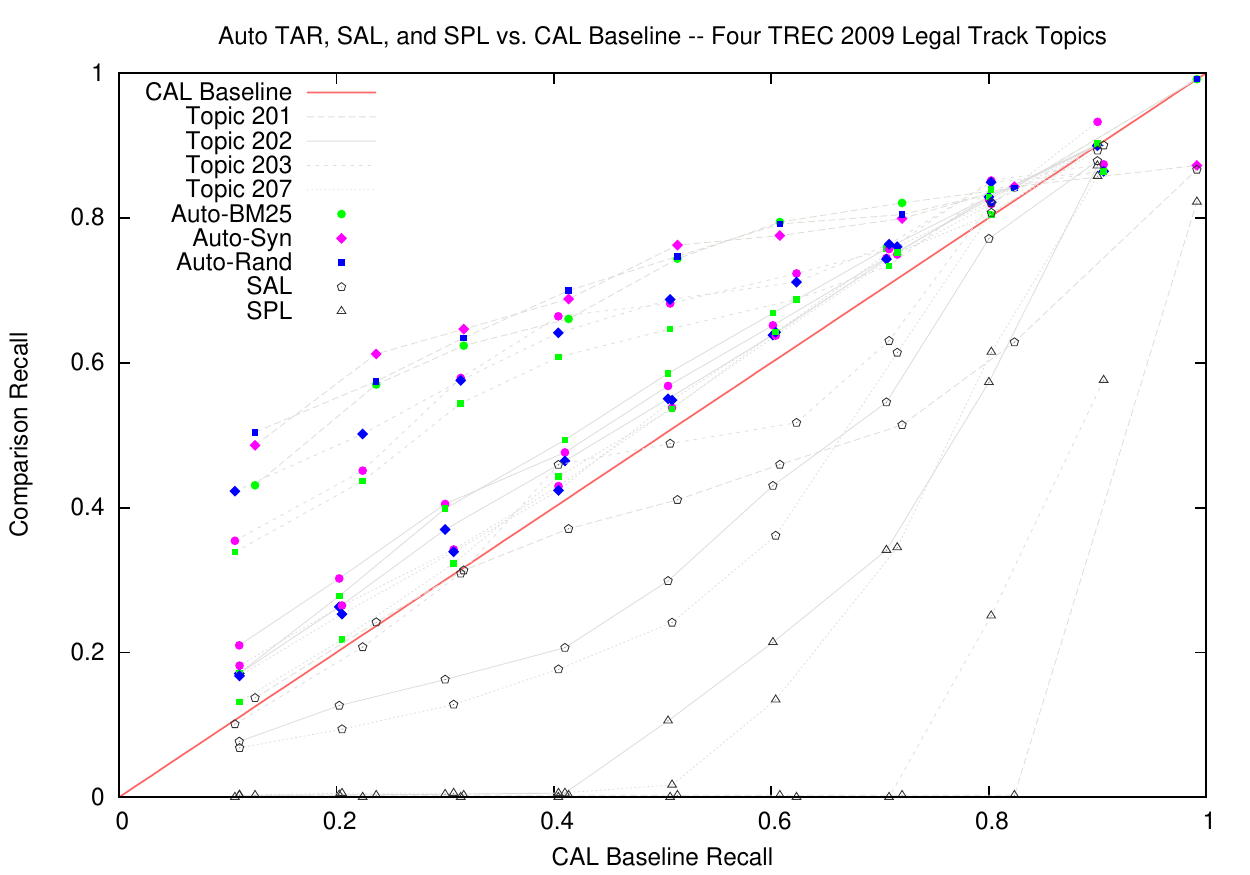}\caption{\label{tab:TREC-2009-Legal}Recall of Auto TAR, and CAL, SAL, and
SPL, using four TREC 2009 Legal Track topics.}
\end{figure*}
The autonomous TAR (``Auto TAR'') process proceeds as follows:
\begin{enumerate}
\item Find a relevant ``seed'' document using ad hoc search, or construct
a synthetic relevant document from the topic description.
\item The initial training set consists of the seed document identified
in step 1, labeled ``relevant.''
\item Set the initial batch size $B$ to $1$.
\item Temporarily augment the training set by adding 100 random documents
from the collection, temporarily labeled ``not relevant.''
\item Train an SVM classifier using the training set.
\item Remove the random documents added in step 4.
\item Select the highest-scoring $B$ documents for review.
\item Review the documents, coding each as ``relevant'' or ``not relevant.''
\item Add the documents to the training set.
\item Increase $B$ by $\left\lceil \frac{B}{10}\right\rceil $.
\item Repeat steps 4 through 10 until a sufficient number of relevant documents
have been reviewed.
\end{enumerate}
Our implementation used a feature space consisting of words occurring
at least twice in the collection, and, following Lewis et al. \cite{lewis2004rnb},
Porter stemming, elimination of SMART stopwords, and Cornell \emph{ltc}
term weighting. SVMlight was used with default parameters. For comparison,
we also modified Cormack and Grossman's CAL, SAL, and SPL implementations
\cite{Cormack:2014:EMP:2600428.2609601} to use the same features.

\subsection{Pilot experiments}

The results of our efforts on the four TREC 2009 Legal Track development
topics are shown in Figure \ref{tab:TREC-2009-Legal}. The left panel
shows the average over the four topics of the gain curves for three
versions of Auto TAR, compared to our modified versions of Cormack
and Grossman's CAL, SAL, and SPL. The three Auto TAR runs differ in
how the seed document was selected: The run labeled ``Auto-BM25''
was seeded with the top-ranked document resulting from Cormack and
Grossman's seed query, using a Wumpus search with BM25 with default
parameters; the run labeled ``Auto-Rand'' was seeded with a random
relevant document; the run labeled ``Auto-Syn'' was seeded using
a synthetic document created using the relevant request for production
(``RFP'') from the mock complaint given to TREC 2009 Legal Track
participants at the outset of the task, with boilerplate language
removed. Table \ref{fig:Seed-queries-and} shows the four seed queries
and the four synthetic seed documents created from the RFPs.

Generally, all Auto TAR runs achieve moderate levels of recall with
less review effort than CAL, but for very high levels of recall are
indistinguishable from CAL. The SAL and SPL gain curves are generally
inferior, consistent with the results reported by Cormack and Grossman
\cite{Cormack:2014:EMP:2600428.2609601}. Averages, such as those
presented in the left panel of Figure \ref{tab:TREC-2009-Legal},
are generally inadequate to assess the reliability of Auto TAR. For
this reason, we recapitulate the same results as a differential plot,
shown in the right panel. Each point in this plot compares the recall
achieved by one method with the recall achieved by another, on the
same topic, for the same review effort. In effect, it stretches the
gain curve for each topic so that the baseline system's results appear
on the diagonal. Points above the diagonal represent improvements
on the baseline; points below a degradation. Light gray lines connect
points representing the same run. The differential plot shows that
the Auto TAR runs, represented by solid, colored points, are superior
to the CAL baseline for almost all combinations of run, topic, and
recall level. The SAL points mostly fall below the baseline, while
the SPL points all fall below the baseline.

\begin{table}
\begin{centering}
\begin{tabular}{|c|c|c|c|c|c}
\cline{1-5} 
\multirow{2}{*}{Topic} & \multirow{2}{*}{CAL} & \multicolumn{3}{c|}{Auto TAR} & \tabularnewline
 &  & \multicolumn{1}{c}{BM25} & \multicolumn{1}{c}{Synth.} & Rand. & \tabularnewline
\cline{1-5} 
201 & 3,400 & 2,400 & 2,200 & 2,400 & \tabularnewline
202 & 9,100 & 8,300 & 8,300 & 8,000 & \tabularnewline
203 & 4,800 & 3,800 & 4,100 & 4,300 & \tabularnewline
207 & 9,400 & 8,200 & 8,000 & 8,000 & \tabularnewline
\cline{1-5} 
\end{tabular}
\par\end{centering}

\caption{\label{tab:CALpilotRE75}75\% Recall effort for CAL and Auto TAR.}
\end{table}
Table \ref{tab:CALpilotRE75} shows 75\% recall effort for our implementation
of CAL compared to the three variants of Auto TAR. From the table,
it is apparent that the Auto TAR runs all require less effort than
CAL to achieve 75\% recall, but there is no clear winner among them.
We note that our CAL implementation achieves somewhat lower recall
effort than the implementation in Cormack and Grossman's Toolkit,
which we attribute to our feature engineering choice (i.e., tf-idf
word-based as opposed to binary byte 4-grams).

\subsection{Hypotheses and design choices}

\subsubsection{Single relevant seed document}

One of our principal design choices, which influenced all others,
was to use a single relevant seed document, instead of the 1,000-document
seed set in Cormack and Grossman's CAL implementation. This decision
was motivated by several factors. Perhaps most important, was our
desire to avoid the situation in which the seed set contained no relevant
examples, and hence no basis for relevance feedback. We were unwilling
to revert to random search because the expected effort to find a relevant
document -- on the order of $\frac{1}{\rho}$ where $\rho$ is the
prevalence of relevant documents in the collection -- would be unacceptable
when prevalence was low. Even in situations where there were one or
several relevant documents in the seed set, we were concerned that
for such a sparse training set, it would be difficult to set regularization
and other parameters for the SVM implementation to converge without
overfitting. One of Cormack and Grosman's examples (Topic 203, CAL
and SAL, with random seed \cite{Cormack:2014:EMP:2600428.2609601})
appears to exhibit this abnormality.

\begin{table}
\begin{tabular}{|r|>{\raggedright}p{0.45\linewidth}|>{\raggedright}p{0.45\linewidth}|}
\hline 
Topic & Seed Query & Synthetic Seed Document\tabularnewline
\hline 
201 & ``pre-pay'' OR ``swap'' & engagement in structured commodity transactions known as prepay transactions\tabularnewline
\hline 
202 & ``FAS'' OR ``transaction'' OR ``swap'' OR ``trust'' OR ``transferor''
OR ``transferee'' & engagement in transactions that Enron characterized as compliant with
FAS 140 (or its predecessor FAS 125)\tabularnewline
\hline 
203 & ``forecast'' OR ``earnings'' OR ``profit'' OR ``quarter''
OR ``balance sheet'' & had met, or could, would, or might meet its financial forecasts, models,
projections, or plans at any time after January 1, 1999\tabularnewline
\hline 
207 & ``football'' or ``Eric Bass'' & fantasy football, gambling on football, and related activities, including
but not limited to, football teams, football players, football games,
football statistics, and football performance. \tabularnewline
\hline 
\end{tabular}\caption{\label{fig:Seed-queries-and}Seed queries and synthetic seed documents.}
\end{table}
Our motivation to use a single relevant seed document also stemmed
in part from criticism of the mechanism to determine its content.
It has been argued that unless the seed set is ``representative of
the collection'' a TAR effort may fail to yield adequate recall,
and that using keywords or other judgmental techniques may ``bias''
the result \cite{schieneman2013implications}. We wished to minimize
and isolate the judgmental input needed to initiate the TAR process,
so as to analyze its impact. To this end, we chose three methods of
selecting the seed document: random, BM25, and synthetic. While we
do not believe that random selection is a generally applicable method,
due to the low-prevalence issue referenced above, it serves as a proxy
for a relevant document already known to the user, or identified by
any number of methods. If a random document nearly always fits the
bill, such a convenience-sampled document should, as well. We had
reason to think that the top-ranked (according to BM25, in our experiments)
relevant document from an ad hoc search might be more effective, or
more reliable, than a random one, but we could also imagine scenarios
in which it might be worse, perhaps resulting in a myopic review.
We further posited that a synthetic document consisting of a description
of the subject matter would make a good seed for the same reason that
such a document would provide a reasonable query for a relevance-ranking
algorithm for ad hoc IR. The use of a synthetic seed document offers
the simplicity of a turnkey approach. The use of a BM25 seed is predicated
on a two-phase approach: an ad hoc search to find a document, followed
by Auto TAR. Generally, ad hoc search will yield a relevant document
in the first several attempts; if not, the failure will be readily
apparent, and the user will reformulate the query rather than reviewing
to an absurd depth in the ranking. In our comparisons, we report only
the effort required for the Auto TAR phase; we leave it to the reader
to account for the effort of finding the seed: For a synthetic seed,
there is no search effort; for a BM25 seed, the search effort is usually
de minimus; for a truly random seed, the effort is on the order of
$\frac{1}{\rho}$; for an arbitrary seed, there is no search effort
provided at least one relevant document is known.

\subsubsection{Presumptive non-relevant examples}

When the seed set is restricted to a single relevant seed document,
we lack non-relevant documents with which to train the SVM, rendering
it unable to find a meaningful decision boundary. Instead of having
the reviewer assess random documents for this purpose, we presumptively
and temporarily label 100 randomly selected documents ``not relevant''
for the purpose of training the SVM. We repeat this procedure -- augmenting
the training set by a different set of 100 randomly selected documents,
presumptively labeled ``not relevant,'' from the documents yet to
be reviewed -- for each iteration of relevance feedback. Our rationale
is as follows: For low prevalence topics ($\rho\ll0.01$), there will
likely be no relevant documents among those presumptively labeled
``not relevant''; for high prevalence topics ($0.01\ll\rho<0.5$),
there will likely be some relevant documents, but even more non-relevant
documents, and it is unlikely that the resulting SVM will be so poorly
trained as to be unable to find sufficient relevant documents to proceed,
given their high prevalence. Moreover, the choice of a different set
of non-relevant examples introduces enough nondeterminism that poor
training is unlikely to persist through several iterations. The intermediate
case of $\rho\approx0.01$ falls between the extremes; we see no reason
why it should fare worse.

We had reason to believe that the nondeterminism introduced by the
use of random presumptively non-relevant examples might aid in the
coverage of diverse aspects of the topics, for much the same reason
that randomization can help hill-climbing methods avoid local optima.
We conducted one auxiliary experiment that supported our impression:
Increasing the size of the set of documents to 1,000 appeared to degrade
performance. We conjecture that the reason is, in part, because such
a large set smooths out the randomness. Another possibility is that
the SVM simply overfits with this large an imbalance between relevant
and non-relevant examples. Because 100 was our intuitive choice and
appeared to work well, we did not investigate other sample sizes.

\subsubsection{Exponential batch sizes}

Instead of using a batch size of 1,000 for relevance feedback as Cormack
and Grossman did, we were inclined to explore the boundary case of
using a batch size of 1; i.e., retraining the SVM and selecting the
single highest-ranked document. This minimal batch size may afford
the process the greatest possible opportunity to learn, and hence
to achieve high precision. On the other hand, it may deprive the algorithm
of sufficient real (as opposed to presumptive) non-relevant examples
to clearly mark the decision boundary \cite{sculleyonesided}. An
exploration of this issue is met with the formidable problem that
the overall running time of such a solution is $\Omega(n^{2})$, where
$n$ is the size of the collection, by virtue of the fact that it
is necessary to re-score the collection (or, at least, those documents
not yet reviewed) $n$ times. Furthermore, if the training time $T(n)$
is superlinear, the lower bound rises to $\Omega(n\cdot T(n))$. Exploration
aside, a method with quadratic time complexity is simply not viable
for TAR.

We sought to reap the benefits of early precision, while avoiding
downside risk and excessive running time, by using exponentially increasing
batch sizes. We used an initial batch size of 1, and increased it
at each step by the smallest number of documents greater than 10\%.
We chose a growth rate of 10\% because it resulted in about the same
number of iterations, and hence similar computation time, as the fixed
batch size of 1,000 used in Cormack and Grossman's Toolkit. It is
easy to show that the asymptotic running time of this approach is
$O(n\log n+T(n))$, assuming that an $O(n\log n)$ sorting algorithm
is used to select the top-ranked documents.

\subsubsection{Feature engineering and learning method}

Our choice of tf-idf word-based feature engineering and SVMlight,
in preference to binary byte 4-grams and Sofia-ML, was partly occasioned
by our concern over sparse training sets. We were unable to find a
feature engineering method and configuration of Sofia-ML (or Vowpal
Wabbit, or an implementation of logistic regression) that worked for
sparse training sets as well as tf-idf features and SVMlight, with
default parameters. A second consideration was the widespread adoption
of tf-idf and SVMlight, facilitating better-controlled comparisons
with prior work. On the other hand, the use of words and stemming
presupposes English documents, and, to our knowledge, the asymptotic
training time $T(n)$ for SVMlight is superlinear, whereas it is linear
for Sofia-ML, Vowpal Wabbit, and other stochastic methods. Nevertheless,
the running time was adequate for the test collections we used.

\section{Experiments}

\subsection{Actual legal matters}

\begin{figure*}
\includegraphics[width=1\columnwidth]{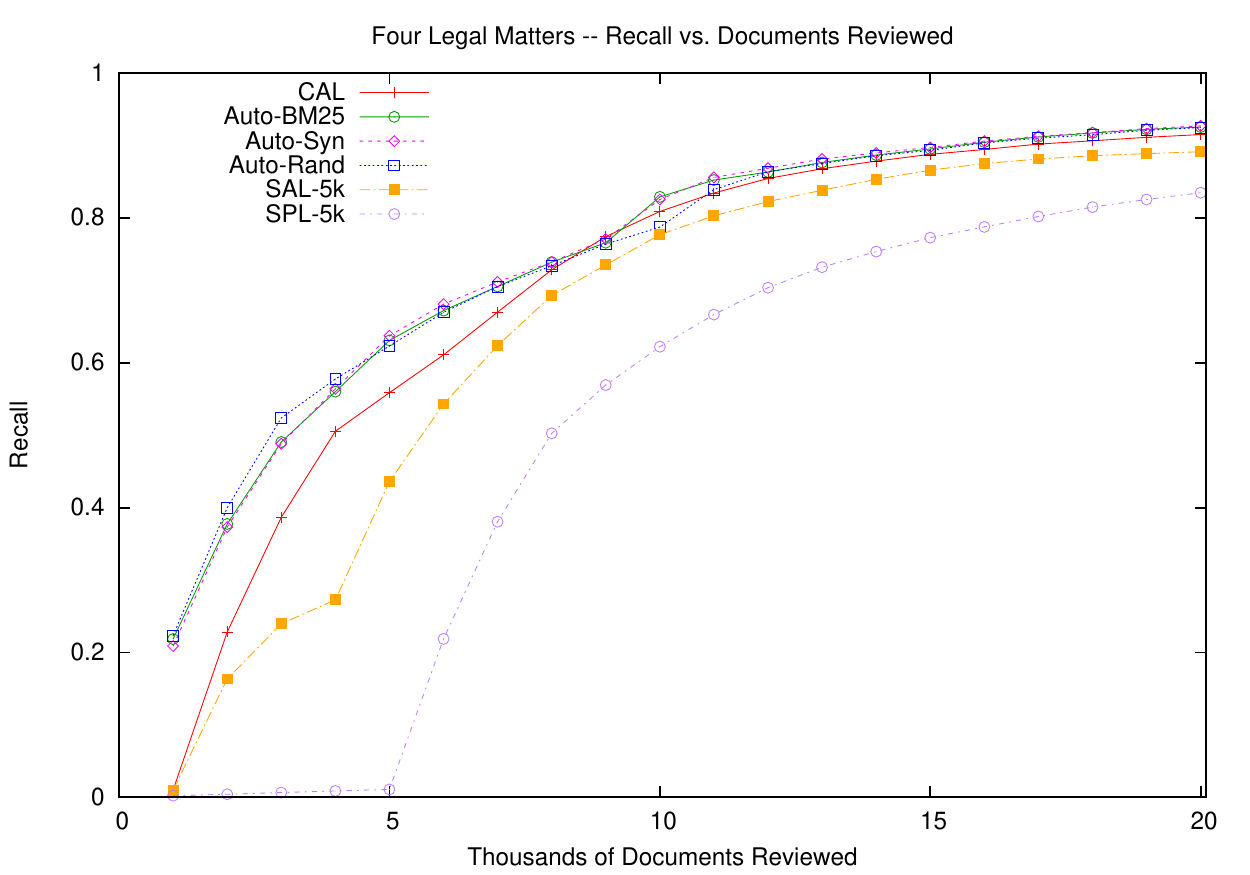}

\includegraphics[width=1\columnwidth]{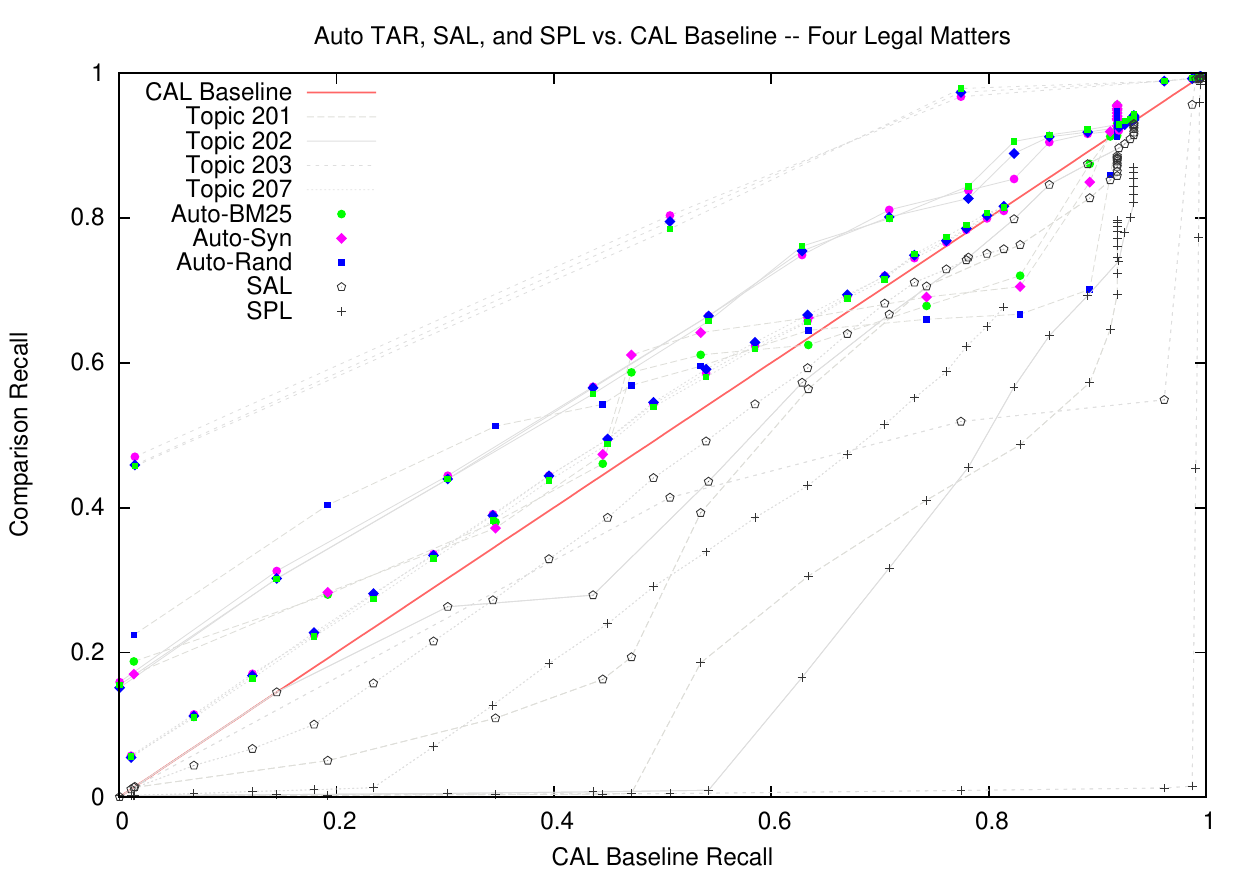}\caption{\label{fig:Four-real-legal}Recall of Auto TAR, and CAL, SAL, and
SPL, using four actual legal matters from Cormack and Grossman.}
\end{figure*}

We arranged to run Auto TAR, configured exactly as described in Section
3.1, on the test collections for the four actual legal matters studied
by Cormack and Grossman \cite{Cormack:2014:EMP:2600428.2609601}.
To this end, it was necessary to create a version of the Toolkit that
imported the raw text, constructed the feature representation, and
ran Auto TAR. The results were captured using the Toolkit. We compared
the three versions of Auto TAR to our reimplementations of CAL, SAL,
and SPL. For SAL and SPL, we chose a representative training set size
of 5,000 documents. The left panel of Figure \ref{fig:Four-real-legal}
shows the average recall over the four topics, while the right panel
shows the differential scatterplot with respect to the CAL baseline.
These results generally mirror those presented in Section 3.2. We
note that for one topic (Matter A), the three AutoTAR runs lag somewhat
for precision values between about 0.65 and 0.85, and then catch up.
We have no explanation for this behavior and would have examined and
reported on the nature of the documents that were retrieved earlier
by CAL than by Auto TAR, had we been able to do so. Lacking that information,
we note that, for this dataset, the simulated reviewer assessments
had a precision of only 0.31, well below that of any other topic.
It may be that Auto TAR was simply finding documents that the reviewer
thought were relevant, but the ultimate authority did not.

\subsection{Reuters RCV1-v2}

\begin{figure*}
\includegraphics[width=1\columnwidth]{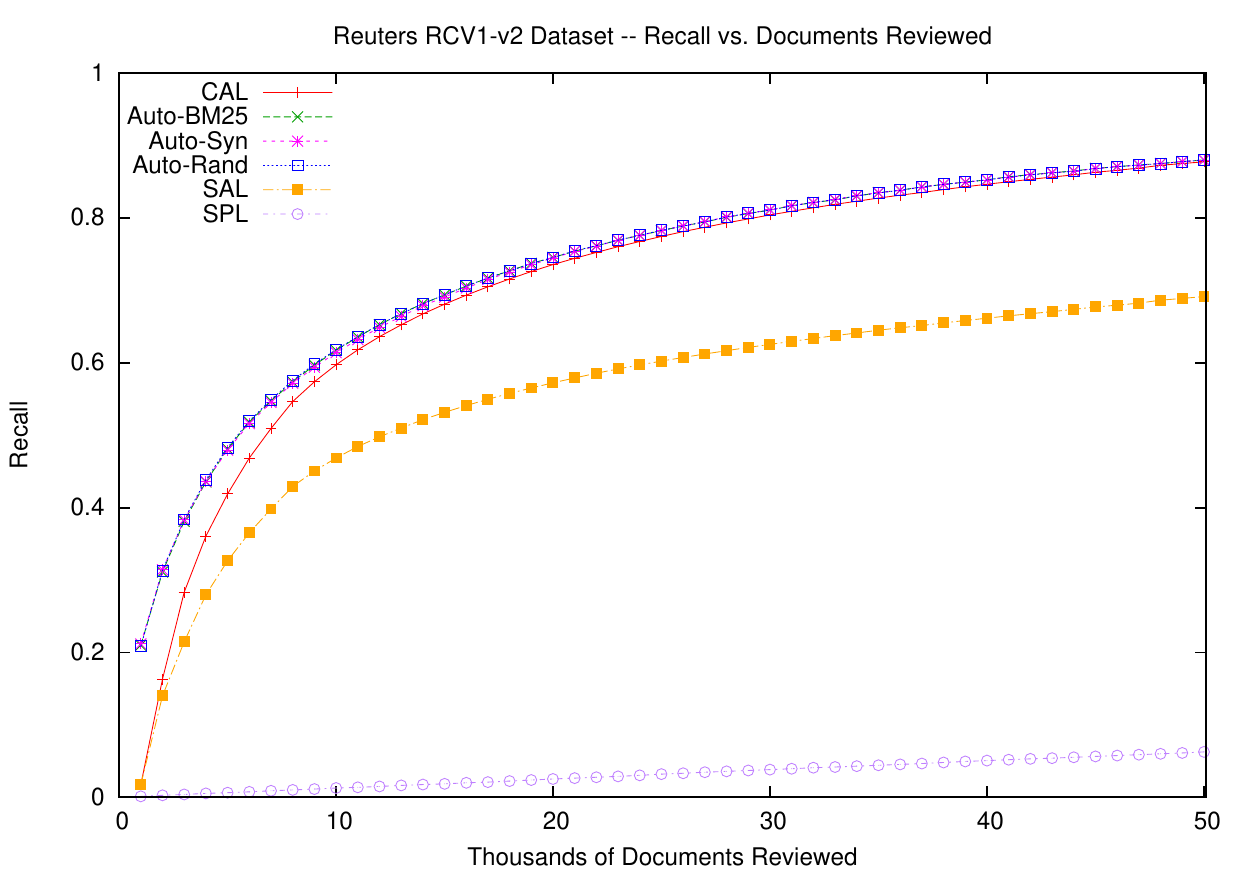}

\includegraphics[width=1\columnwidth]{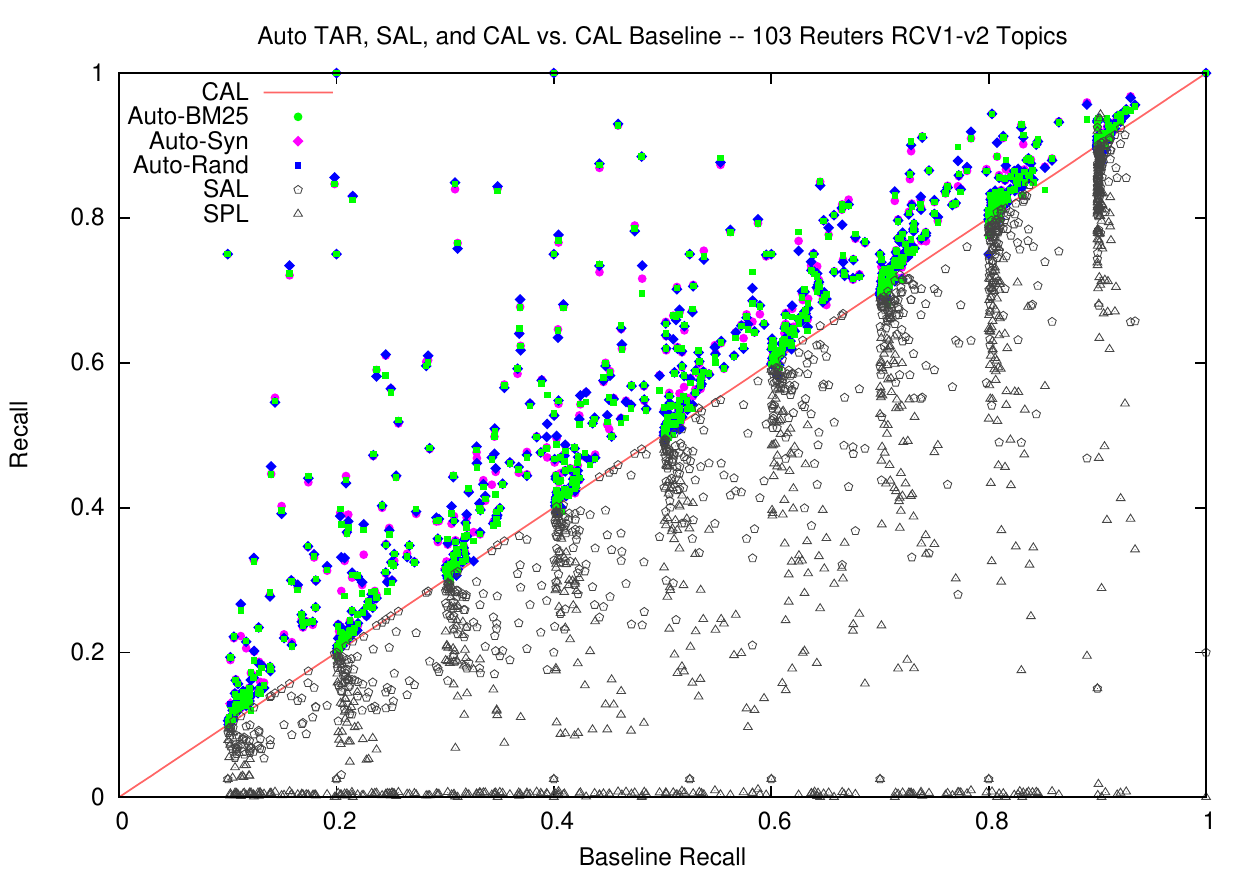}\caption{\label{fig:Subject-Category-Retrieval:}Recall of Auto TAR, and CAL,
SAL, and SPL, using 103 subject-matter topics from Reuters RCV1-v2.}
\end{figure*}
\begin{figure*}
\includegraphics[width=1\columnwidth]{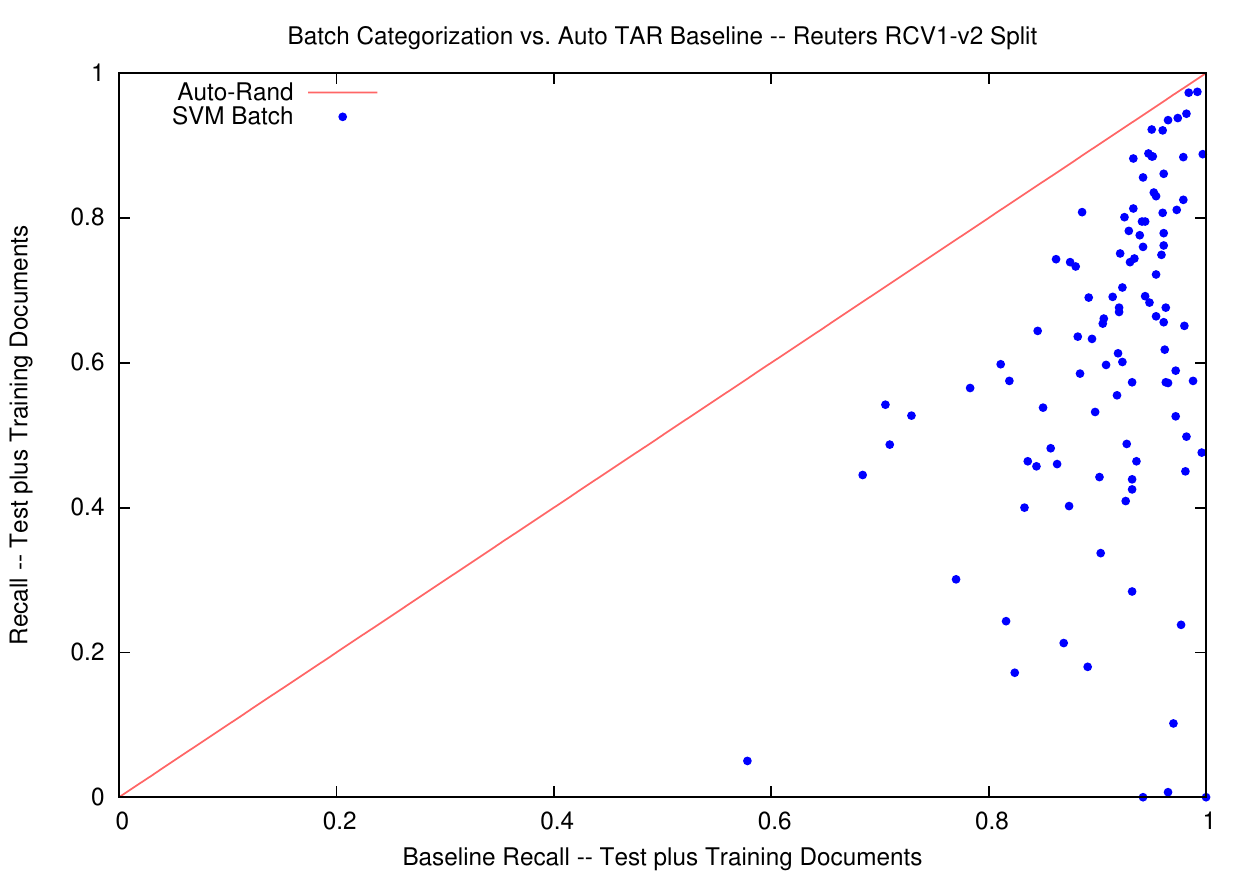}

\includegraphics[width=1\columnwidth]{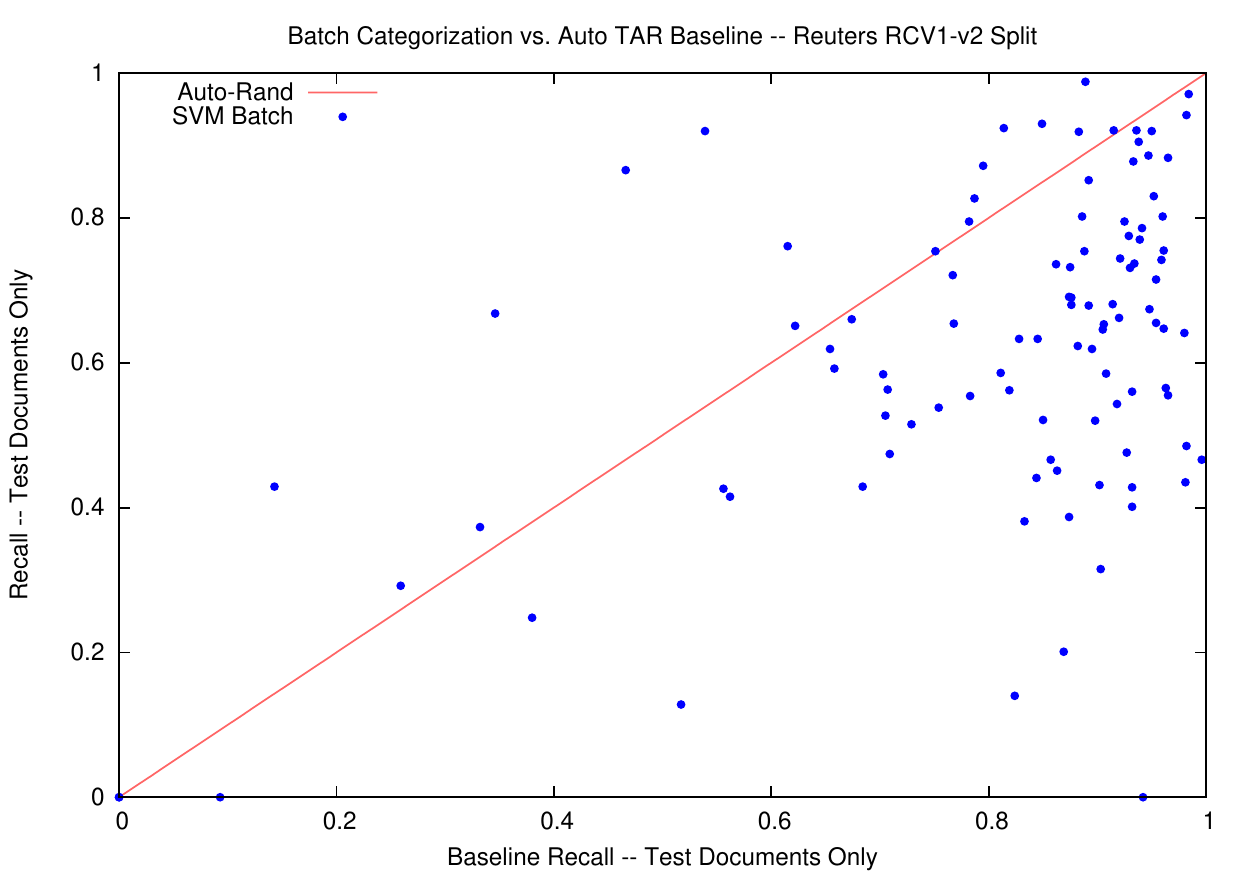}\caption{\label{fig:Text-Categorization:-Reuters}Recall of Auto TAR vs. RCV1-v2
Text Categorization ($F_{1}$ optimized).}
\end{figure*}
The Reuters RCV1-v2 corpus, compiled by Lewis et al. \cite{lewis2004rnb},
consists of 804,401 documents, 23,149 of which are denoted training
documents, and the rest, test documents. Complete labels are available
for each of 103 subjects, 364 industries, and 326 geographic locations.
We ignored the training/test split and used the entire document set
in our experiment. While the subjects form a hierarchy, we treated
each as a separate topic. We used the supplied labels for both training
and evaluation; that is, the ``training standard'' and ``gold standard''
in the Toolkit were identical. For the seed query, we used the subject
header in the RCV1-v2 table of contents, e.g., ``Consumer Prices.''
For the synthetic seed document, we used the subject description,
also supplied with the dataset, along with the subject descriptions
above it in the hierarchy, e.g., ``consumer prices and price indices;
inflation, prices and price indices; ALL Economics and Economic Indicators.''

Our feature engineering choices are virtually identical to those used
for the feature representation supplied with the collection, but we
applied our own, for a number of reasons. First, we did not want to
violate our autonomy constraint of doing no dataset-specific configuration.
Second, the supplied features excluded documents not appearing in
the test set, with the effect that words occurring fewer than 40 times
had a high probability of being excluded, potentially affecting Auto
TAR's ability to harness rare but informative terms. Finally, we were
not able to replicate the version of the Porter stemmer used to create
the supplied features, and were therefore unable to featurize our
synthetic seed documents.

To validate our feature engineering, we computed macro-averaged $F_{1}$
using using SVMlight and the RCV1-v2 splits. Using the supplied feature
representation, we achieved $F_{1}=0.607$, exactly as reported by
Lewis et al. \cite{lewis2004rnb} for the same approach. Using our
features, we achieved $F_{1}=0.608$.

Our first experiment replicates the pilot experiments of Section 3.2,
comparing three versions of Auto TAR with CAL, SAL, and SPL. The left
panel of Figure \ref{fig:Subject-Category-Retrieval:} show that,
on average, the three versions of Auto TAR are indistinguishable,
and superior to CAL, while SAL and SPL are inferior. The right panel
shows that for virtually all combinations of topic and recall, all
versions of Auto TAR are superior or equal to CAL, while SAL and SPL
are equal or inferior.

We conducted a supplemental experiment to compare the effectiveness
of Auto TAR with the supervised text categorization results we achieved
using the RCV1-v2 splits. As noted above, our classification results
were virtually indistinguishable from those reported by Lewis et al.,
which, we understand, still reflect the state of the art. Because
TAR and text categorization are different tasks, we must decide how
to account for the effort involved in training the classifier. Figure
\ref{fig:Text-Categorization:-Reuters} shows the result under two
possible interpretations. The left panel shows the recall of the categorization
with respect to the baseline of Auto TAR with a random seed document,
where the training documents are included in both the recall and effort
calculations. This corresponds to an implementation of SPL where the
cost (and benefit) of training is attributed to the TAR effort. The
right panel shows effort and recall when the training documents are
excluded from the calculation. This corresponds to the situation where
a labeled training set is available as a sunk cost, but it is still
necessary to review documents from the test set. Under the SPL model,
all the categorization results are inferior to the baseline. Perhaps
surprisingly, the vast majority are also inferior under the sunk-cost
model.
\begin{figure*}
\includegraphics[width=1\columnwidth]{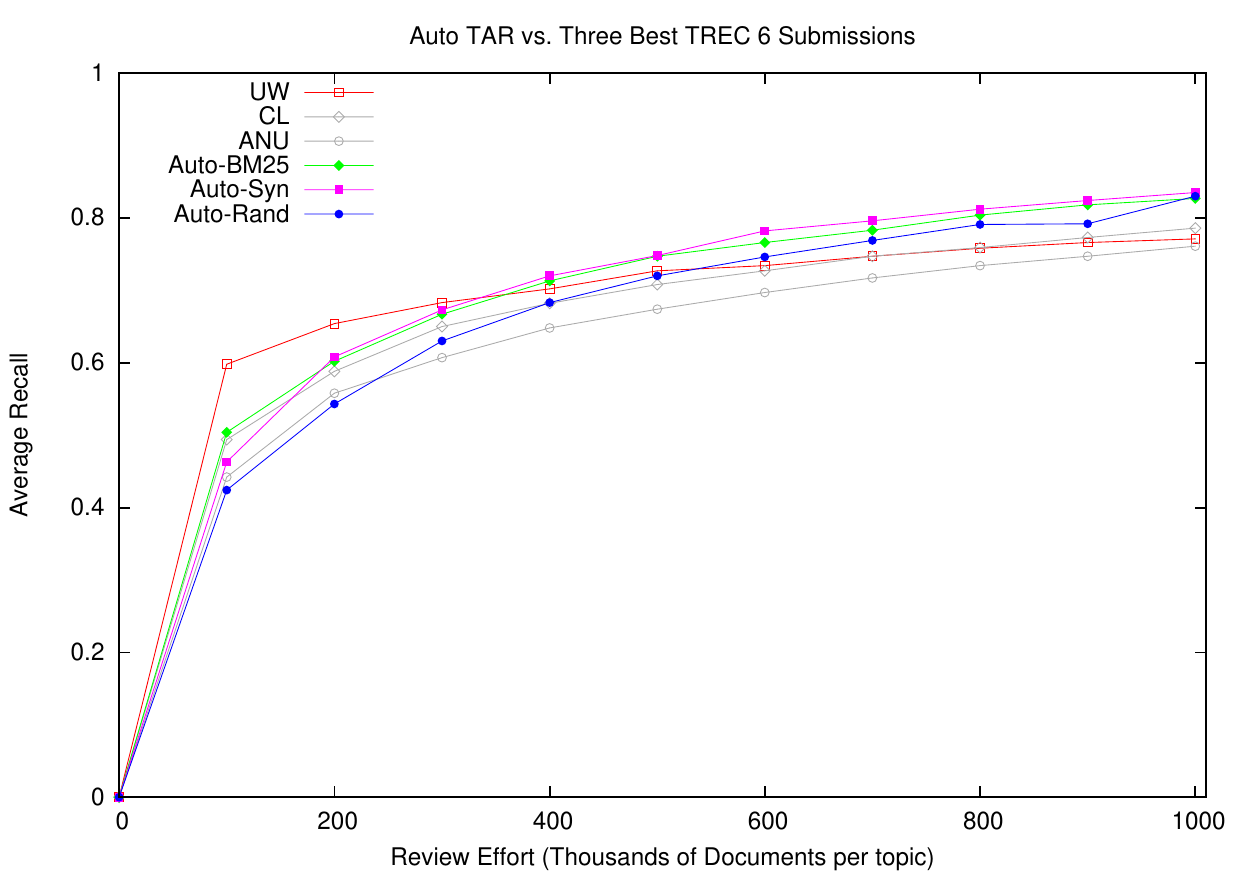}

\includegraphics[width=1\columnwidth]{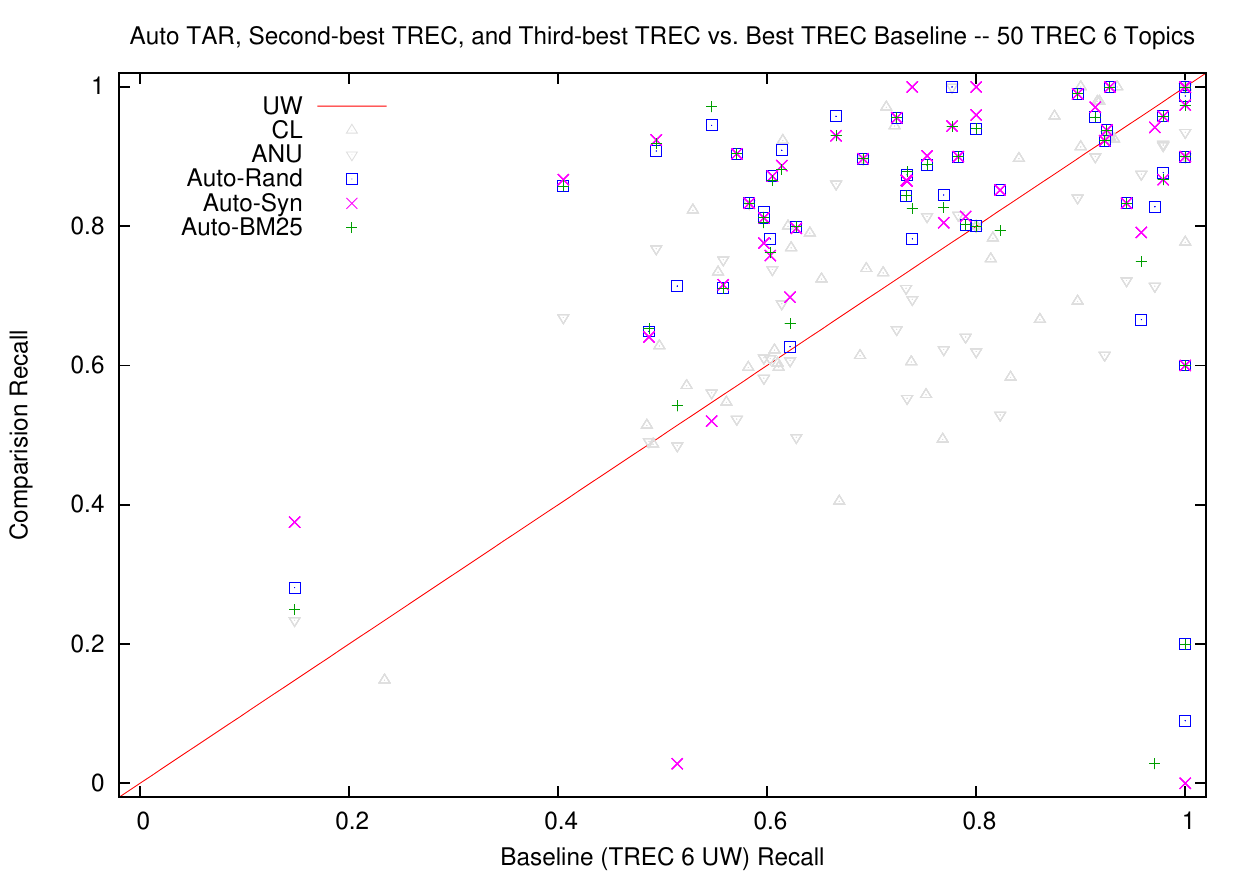}\caption{\label{fig:TREC-6-Ad}Recall of Auto TAR and three best submissions
to TREC 6 Ad Hoc Task.}
\end{figure*}

\subsection{\label{sub:TREC-6-AdHoc}TREC 6 AdHoc Task}

Figure \ref{fig:TREC-6-Ad} compares our three Auto TAR methods to
the Waterloo ISJ effort (``UW'') described in Section 2. The left
panel shows average recall over 50 topics as a function of review
effort per topic. For comparison, two other high-scoring TREC submissions
are shown (denoted as ``LC'' and ``ANU''). While Auto TAR achieves
higher overall recall than the TREC submissions, it appears that UW
reaches 60\% recall more quickly. We note that the Waterloo team re-ranked
their submission to put the most-likely relevant documents first,
so the order shown is not the order in which the documents were retrieved
and reviewed. Therefore, the curve is misleading; one cannot infer
from it that Waterloo achieved an average recall of 0.6 while reviewing
fewer than 100 documents per topic. The scatterplot further reveals
that, for the vast majority of topics, Auto TAR achieves all recall
levels (including $recall=0.6$) with less effort than UW. The average
is apparently the result of a few outlier topics with few relevant
documents, which Waterloo found, and Auto TAR did not.
\begin{figure*}
\includegraphics[width=1\columnwidth]{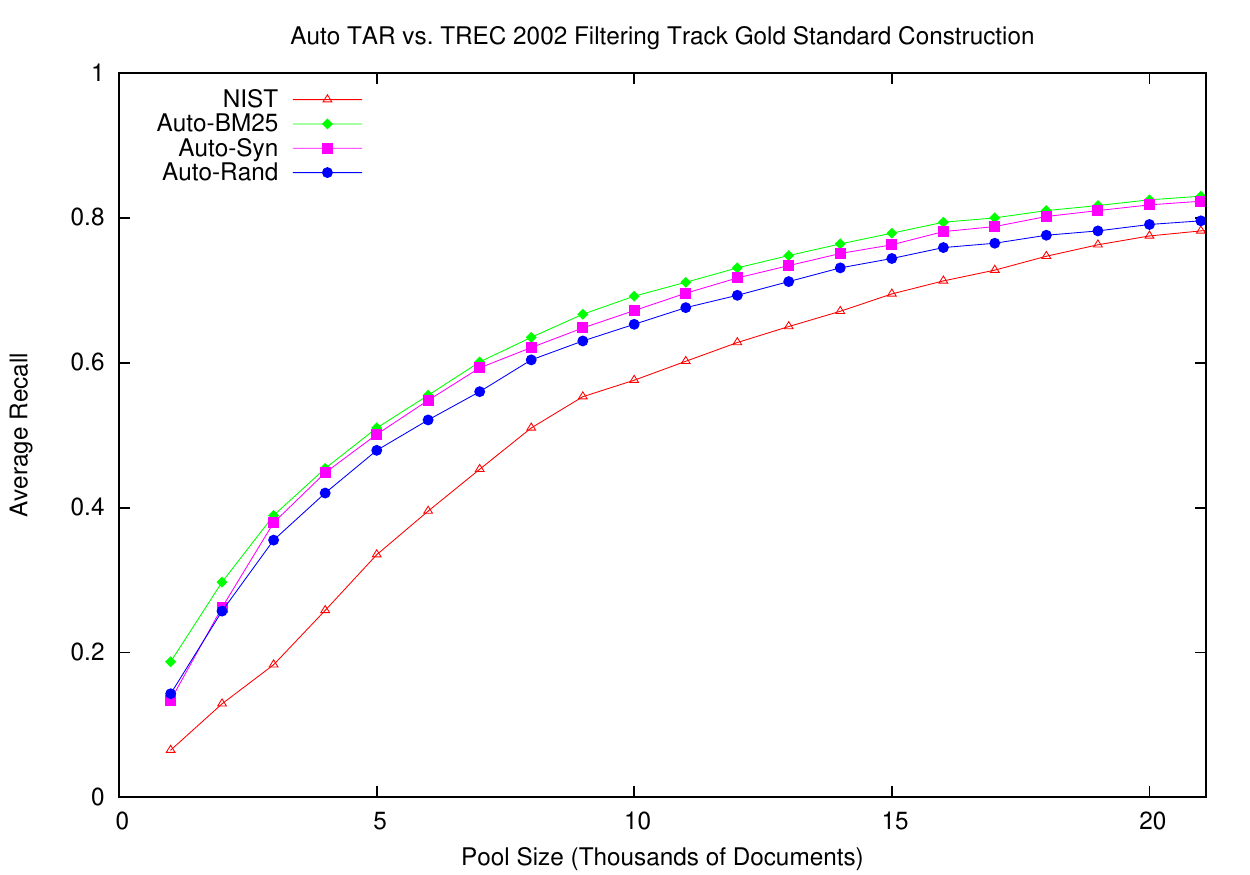}

\includegraphics[width=1\columnwidth]{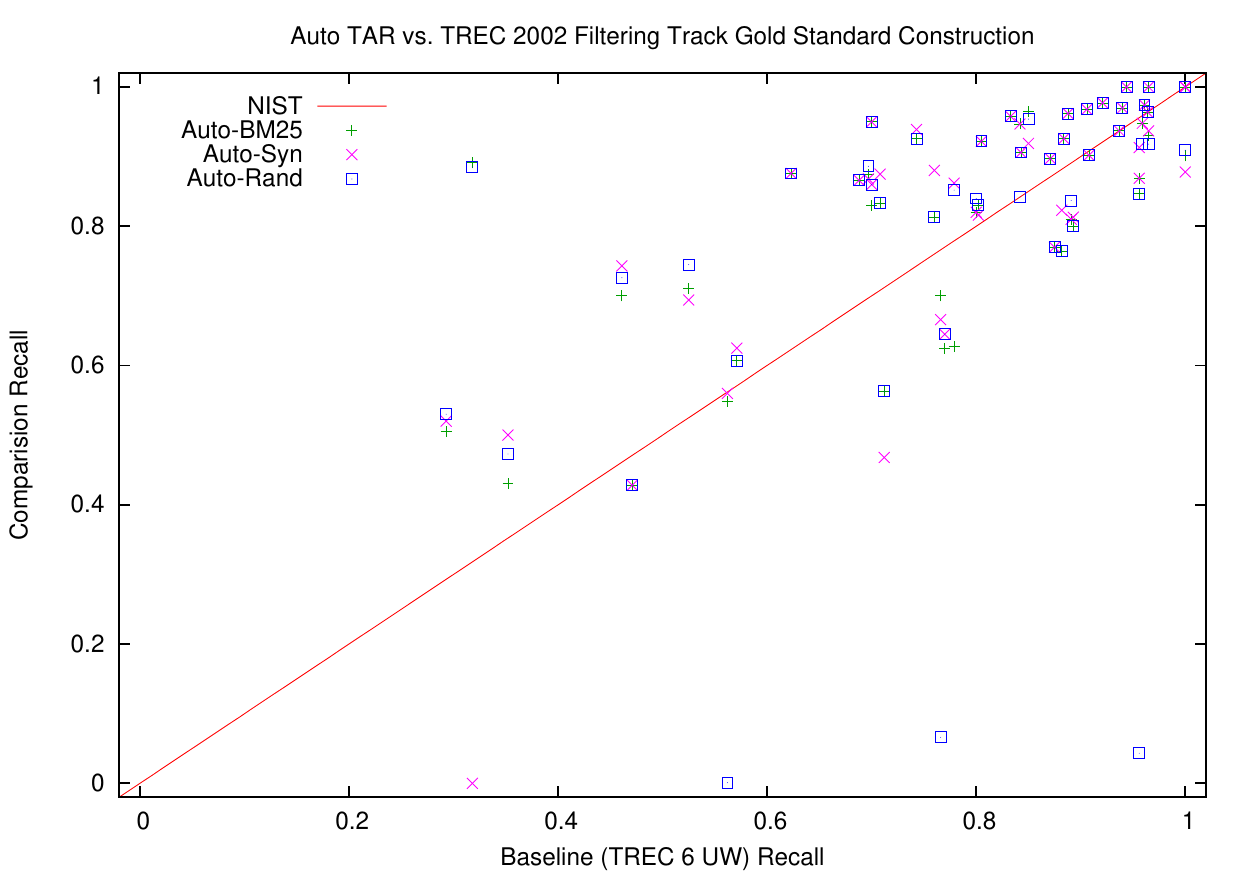}

\centering{}\caption{\label{fig:Construction-of-TREC}Recall of Auto TAR and NIST IRF for
TREC 2002 Filtering Track qrels.}
\end{figure*}
\begin{figure}
\centering{}\includegraphics[width=1\columnwidth]{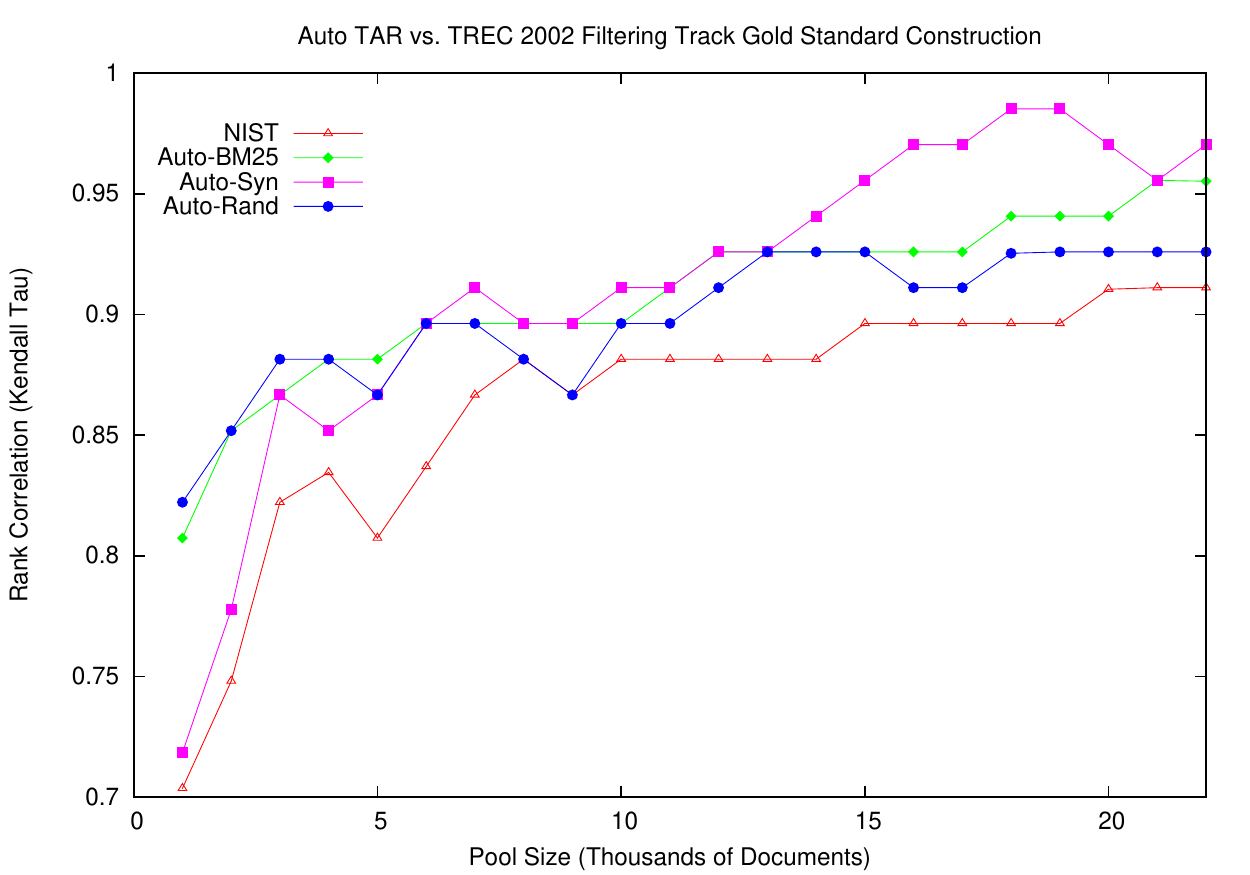}\caption{\label{fig:Rank-Correlation-vs.TREC}Rank correlation of Auto TAR
and NIST IRF qrels with respect to official TREC 2002 Filtering Track
qrels.}
\end{figure}

\subsection{TREC 2002 Filtering Test Collection}

Figure \ref{fig:Construction-of-TREC} shows the average recall of
Auto TAR versus Robertson and Sanderson's work to construct the test
collection for the TREC 2002 Filtering Track (``NIST''), described
in Section 2. The panels repeat the same story as Subsection \ref{sub:TREC-6-AdHoc},
including the presence of a few cases where Auto TAR failed to get
traction. We note that when Auto TAR fails, it seems to do so spectacularly,
retrieving a near-vacuous number of documents. Such a situation would
be readily apparent, and would, we suggest, signal to the user to
abandon the effort and start again with another seed document, or
another method.

Figure \ref{fig:Rank-Correlation-vs.TREC} shows, instead of recall,
the Kendall rank correlation versus the official augmented qrels,
reflecting the effectiveness of the review efforts for their intended
purpose: deriving a good set of qrels.

\section{Conclusion}

Our experiments show that the efficacy of CAL extends well beyond
the eight legal matters studied by Cormack and Grossman \cite{Cormack:2014:EMP:2600428.2609601}.
A handful of principled improvements -- tf-idf features, a single
relevant seed document, presumptively labeled ``not relevant'' examples,
and exponential batch sizes, in combination -- improve recall, especially
at lower effort levels, and almost always improve, not only on CAL
(which is reaffirmed to be consistently superior to SAL and SPL),
but on the best reported efforts for interactive search and judging,
as well as interactive relevance feedback. While a sign test shows
the win rate to be significantly better than chance, ($P\approx0.000$),
our overarching objective is not to win most of the time, but to win
all of the time, or at least not to lose by a substantial margin.
To this end, we offer our qualitative observation that the losses
incurred by Auto TAR generally involved topics for which it had difficulty
finding a non-trivial number of relevant documents beyond the single
seed document. It remains an open question how to enhance Auto TAR
to detect this eventuality and, perhaps, to request a new seed.

Our results indicate that there is little difference, if any, beyond
chance in choosing between a seed selected randomly, a seed selected
by chance, and a synthetic seed constructed from the topic description.
Auxiliary experiments indicated that chance variation between runs
(due to the selection of the seed document, as well as the selection
of the presumptively non-relevant training examples) was much larger
than any systematic difference between seeding methods.

A commonly expressed sentiment in eDiscovery is that there can be
no ``one size fits all'' TAR solution, suggesting that it is necessary
to select tools and strategy with knowledge of the topic and dataset,
and that some tools are more appropriate in some situations. Thus
far, however, we have been unable to find a situation in which one
could choose, without foreknowledge, a tool or strategy to yield better
recall, with less effort, than Auto TAR.

Beyond enhancing the effectiveness and reliability of Auto TAR, it
remains an open question how best to decide when to terminate the
TAR process. The gain curves show clearly diminishing returns at some
point, but do not show exactly how many more relevant documents remain
to be found. Our results indicate that if a substantial number of
relevant documents are found with high precision, and then precision
drops, the vast majority of relevant documents have likely been found.
Methods for achieving a reliable, efficient, quantitative estimate
remain elusive.

Auto TAR demonstrates that reasonable and reliable results can be
achieved without discretionary input to the TAR process, thereby avoiding
the risk of negative bias. It may be that Auto TAR, by providing a
floor level of effectiveness, can still harness discretionary input
to advantage, while avoiding the downside risk. For example, the user
might provide additional seed documents or queries, either when Auto
TAR appears to be stuck in a ditch, or to ensure that all aspects
of relevance are covered, if it could be known that Auto TAR would
achieve at least its floor level of effectiveness.

Our adaptation of Cormack and Grossman's Toolkit, the feature representation
of the TREC and RCV1-v2 collections, as well as gain curves and result
tables for the 211 individual topics, are available in the on-line
appendix. We believe that Auto TAR sets a new standard to beat.

\bibliographystyle{abbrv}
\bibliography{irate}


\end{document}